\documentclass[sigconf]{acmart}
\AtBeginDocument{%
  }

\setcopyright{acmlicensed}

\begin{document}

\title{A Survey of Challenges and Sensing Technologies in Autonomous Retail Systems}

\author{Shimmy Rukundo}
\email{rsnhimiy@umich.edu}
\orcid{1234-5678-9012}
\affiliation{%
  \institution{University of Michigan}
  \city{Ann Arbor}
  \state{MI}
  \country{USA}
}

\author{David Wang}
\email{davwan@umich.edu}
\orcid{1234-5678-9012}
\affiliation{%
  \institution{University of Michigan}
  \city{Ann Arbor}
  \state{MI}
  \country{USA}
}

\author{Front Wongnonthawitthaya}
\email{frontw@umich.edu}
\orcid{734-604-1819}
\affiliation{%
  \institution{University of Michigan}
  \city{Ann Arbor}
  \state{MI}
  \country{USA}
}

\author{Youssouf Sidibé}
\email{sidibey@umich.edu}
\affiliation{%
  \institution{University of Michigan}
  \city{Ann Arbor}
  \state{MI}
  \country{USA}
}

\author{Minsik Kim}
\email{minsikky@umich.edu}
\affiliation{%
  \institution{University of Michigan}
  \city{Ann Arbor}
  \state{MI}
  \country{USA}
}

\author{Emily Su}
\email{syuchien@umich.edu}
\affiliation{%
  \institution{University of Michigan}
  \city{Ann Arbor}
  \state{MI}
  \country{USA}
}

\author{Jiale Zhang}
\affiliation{%
  \institution{University of Michigan}
  \city{Ann Arbor}
  \state{MI}
  \country{USA}}
\email{jiale@umich.edu}

\renewcommand{\shortauthors}{Trovato et al.}

\begin{abstract}
Autonomous stores leverage advanced sensing technologies to enable cashier-less shopping, real-time inventory tracking, and seamless customer interactions. However, these systems face significant challenges, including occlusion in vision-based tracking, scalability of sensor deployment, theft prevention, and real-time data processing. To address these issues, researchers have explored multi-modal sensing approaches, integrating computer vision, RFID, weight sensing, vibration-based detection, and LiDAR to enhance accuracy and efficiency. This survey provides a comprehensive review of sensing technologies used in autonomous retail environments, highlighting their strengths, limitations, and integration strategies. We categorize existing solutions across inventory tracking, environmental monitoring, people-tracking, and theft detection, discussing key challenges and emerging trends. Finally, we outline future directions for scalable, cost-efficient, and privacy-conscious autonomous store systems.
\end{abstract}




\maketitle

\section{Introduction}
Autonomous systems are transforming industries such as warehousing, manufacturing, and retail, enhancing efficiency and reducing operational costs. Warehouses now rely on robotic picking, autonomous guided vehicles (AGVs), and AI-driven inventory tracking to optimize logistics with minimal human intervention. Similarly, manufacturing integrates smart factories, IoT-connected machinery, and predictive maintenance to streamline production and reduce downtime. These advancements are driven by AI, real-time data processing, and sensor fusion, enabling seamless automation.

In retail, autonomous stores leverage computer vision, RFID, weight sensing, and multimodal data fusion to track inventory, monitor customer interactions, and enable cashier-less shopping. Implementations like Amazon Go use AI-powered tracking to eliminate traditional checkout, but challenges remain, including accurate inventory tracking, real-time product detection, theft prevention, and efficient environmental control.

To tackle these issues, researchers have explored various sensing and machine learning techniques. Vision-based tracking faces occlusions and product similarity issues, while weight sensors struggle with granularity for small items. Vibration-based sensing aids in shelf interaction detection and physiological monitoring, whereas RFID enhances product tracking but is costly for large-scale deployment. Additionally, people-tracking is crucial for theft detection and store optimization, with LiDAR, vibration, and RFID-based methods offering privacy-friendly alternatives to camera surveillance.

This survey examines recent advancements in inventory tracking, environmental monitoring, customer interaction detection, and theft prevention. We analyze various sensing modalities, examine their effectiveness, and discuss future directions in scalable and cost-efficient autonomous store technology.

\section{Challenges}
Autonomous stores aim to streamline operations by eliminating manual checkout, improving inventory tracking, and optimizing store management through sensor-driven automation. However, these systems face several technical challenges across multiple domains. Inventory monitoring remains a critical issue, as traditional vision-based tracking struggles with occlusions, similar-looking products, and dynamic customer interactions, while weight and RFID-based methods face scalability and cost constraints. Environmental conditions, such as temperature fluctuations, require precise sensor-based regulation to maintain perishable goods. Additionally, people-tracking is essential for security, theft prevention, and store layout optimization, yet existing solutions must balance accuracy, privacy concerns, and computational efficiency. Furthermore, interaction tracking—determining how customers engage with products—relies on a combination of vibration, RFID, and multimodal sensing to detect pickups, replacements, and unintended product movements. Finally, real-time operation requires fast, synchronized data processing across multiple sensing modalities, ensuring seamless inventory management, security enforcement, and customer experience in fully autonomous retail environments.

\subsection{Inventory Tracking}

Accurate inventory tracking is a fundamental challenge in autonomous stores, as real-time monitoring of product availability directly impacts operational efficiency, customer satisfaction, and revenue.

Vision-based systems are commonly employed for automated inventory tracking. Including in warehouses, and for stores~\cite{kotheeranurak2023long,falcao2021piwims,falcao2021isacs}. These approaches face several limitations. Occlusions present a major challenge, as products may be hidden by other items, customer hands, or store layouts, making real-time recognition difficult. Additionally, similar-looking products, such as different flavors of chips or soft drinks with identical packaging, often confuse vision-based recognition models, reducing accuracy \cite{haritaoglu2001tracking}. While deep learning-based object detection \cite{yumang2022recognition, babila2022object, kanjalkar2023intelligent, iyer2023retail, bharadi2023real, ra2024dynamic} can help mitigate these issues, these models are computationally expensive, requiring significant processing power. To improve efficiency, various studies \cite{kim2023low, pestana2021full, wang2021fpga, fan2018real, kim2023lightweight} have explored hardware acceleration techniques to speed up deep learning inference. However, scalability remains a challenge, as these models still require large, labeled datasets for training, increasing both data collection costs and deployment complexity \cite{ansari2022expert}.

To address the shortcomings of vision-based tracking, multi-modal sensing approaches combine weight, vision, and vibration-based sensing for more accurate inventory monitoring \cite{ruiz2019aim3s, falcao2020faim, guinjoan2023multi, lindermayr2020multi, patel2025multi, liu2024digital}.
While these multi-modal methods improve detection accuracy, they require precise synchronization and high computational power, leading to increased deployment and maintenance costs. Moreover, factors such as sensor calibration drift, noise from customer movements, and real-time processing demands pose additional challenges. Despite these difficulties, multi-modal approaches remain the most promising direction for scalable and cost-effective inventory tracking in autonomous retail environments.

\subsection{Temperature Control}
Maintaining optimal temperature conditions is critical for autonomous stores, especially those handling perishable goods such as fresh produce, dairy, and frozen foods. Ineffective temperature control can lead to spoilage, financial losses, and potential health hazards for consumers. Traditional refrigeration systems rely on thermostat-based regulation, but autonomous stores require more advanced, sensor-driven approaches to ensure real-time monitoring and adaptive control.

\subsubsection{Sensor-Based Temperature Monitoring}

Advanced temperature sensors are employed to monitor and regulate the temperature of perishable goods continuously. These sensors are integrated into an IoT-based architecture, allowing for real-time data collection, processing, and analysis to ensure optimal temperature conditions. Autonomous stores typically use wireless, low-power temperature sensors, such as MEMS-based sensors and resistance temperature detectors, to measure ambient temperature. These sensors communicate data using low-power protocols like LoRaWAN or ZigBee, enabling long-range and low-latency transmission. This allows the system to monitor temperature conditions across the entire store efficiently.

For enhanced accuracy, more specialized sensors are utilized, including thermocouples and infrared sensors. Thermocouples offer broad temperature ranges with high precision, while infrared sensors enable non-contact temperature measurements, which are particularly useful for monitoring refrigerated goods. Additionally, capacitive humidity sensors are used to measure relative humidity, a critical factor affecting the shelf life of moisture-sensitive products. Ultrasonic sensors also monitor humidity levels by detecting variations in sound wave speed caused by moisture changes. Furthermore, RFID-based monitoring systems with embedded temperature sensors are deployed to track goods throughout the supply chain, ensuring real-time logging and transmission of temperature data as items move from warehouses to store shelves.

\subsubsection{Automated Temperature Regulation}

Effective temperature control systems in autonomous stores combine sensor data with intelligent systems for adaptive regulation, ensuring energy efficiency while protecting perishable goods. Automated cooling systems are designed to adjust dynamically to changing environmental conditions, often using compressor-based refrigeration units with variable-speed fans and evaporators. By modulating compressor speed and fan intensity based on real-time sensor inputs, these systems maintain precise temperature control while minimizing energy consumption. The integration of machine learning models enhances the efficiency of cooling systems by predicting potential system failures and optimizing cooling strategies. These systems rely on supervised learning techniques, such as regression models or more advanced time-series forecasting models, to analyze historical temperature data and environmental conditions, adjusting operations to maintain optimal store climates and reduce energy waste.

In addition to automated cooling, AI-driven predictive maintenance plays a crucial role in maintaining the reliability of refrigeration systems. Predictive maintenance models analyze historical data and current sensor inputs to anticipate potential failures, allowing for early intervention. For instance, machine learning algorithms can forecast temperature fluctuations or system malfunctions before they occur, triggering preventive maintenance actions or adjusting refrigeration setpoints proactively. Adaptive control algorithms adjust cooling intensity based on real-time factors like customer density and external weather conditions, optimizing energy use and temperature stability. Cold chain logistics also benefit from IoT-enabled temperature tracking, with GPS systems and real-time temperature logging devices ensuring that refrigerated goods are transported under optimal conditions and deviations are promptly addressed to prevent spoilage.

Future advancements in temperature control for autonomous stores will focus on AI-driven predictive cooling, energy-efficient refrigeration, and self-healing sensor networks to improve reliability and reduce operational costs. Advanced technologies, such as solid-state cooling and quantum dot-based thermal regulation, could offer more precise and sustainable solutions. However, key challenges remain, including sensor calibration drift, network latency in real-time monitoring, and high deployment costs for advanced systems. Additionally, integrating multi-modal sensing with existing store infrastructure requires careful synchronization to prevent false alarms and inefficiencies. Addressing these challenges will be crucial for scalable, autonomous temperature management in future retail environments.

\subsection{People-Tracking}
 Tracking people in autonomous retail environments is crucial for optimizing store layouts, enhancing loss prevention, and understanding customer behavior. Unlike traditional camera-based surveillance systems, which pose privacy risks and require high computational power, modern approaches leverage multimodal sensing to achieve accurate and privacy-preserving tracking. Methods such as LiDAR, RFID, vibration sensing, and weight-based tracking offer scalable and efficient alternatives. While vision-only approaches have been prevalent, this paper focuses on other sensing types.

Accurate people-tracking in retail stores faces several challenges:
\begin{itemize}
    \item Occlusion \& Crowd Density: Multiple customers entering together or interacting near shelves can block each other from sensor detection.
    \item Privacy Concerns: AI-based camera surveillance raises ethical and legal issues, leading to growing restrictions under GDPR and CCPA regulations.
    \item Differentiating People from Objects: Carts, strollers, or merchandise can interfere with motion-based tracking systems.
    \item Group Shopping Behavior: A family or group may enter together, but only one person makes a purchase, making foot traffic alone an incomplete metric for sales analysis.
\end{itemize}
To address these challenges, multimodal sensor fusion combines multiple sensing techniques to improve detection accuracy without relying solely on cameras.

\subparagraph{\textbf{LiDAR for Depth-Based Tracking}}
LiDAR sensors emit laser pulses to measure depth, creating a 3D map of a store’s entrance or aisles. Unlike cameras, LiDAR does not capture facial details, making it privacy-preserving while still being accurate enough to count customers.

Studies such as FAIM \cite{falcao2020faim} have explored combining weight-based tracking with vision to differentiate customer movements, reducing errors caused by occlusion. However, replacing vision-based methods with LiDAR-only solutions could offer a privacy-first alternative while still ensuring accuracy.

\subparagraph{\textbf{RFID for Shopper Identification}}
RFID (Radio Frequency Identification) is widely used in autonomous shopping to track product movement. When integrated with entry/exit tracking, RFID-enabled loyalty cards or wearable tags can associate individuals with purchases, solving issues related to group entry scenarios where only one person in a group buys an item.

For example, RFID exit detection systems \cite{bocanegra2020rfgo} scan tags at store exits, ensuring all products are accounted for before a customer leaves. However, RFID-based tracking alone cannot differentiate multiple individuals, making it less reliable for detailed people-tracking without additional sensors.

\subparagraph{\textbf{Vibration-Based Tracking}}
Floor vibrations can provide passive, unobtrusive tracking of people’s movement patterns~\cite{pan2017footprintid,pan2015indoor,dong2022stranger,mirshekari2021obstruction,mirshekari2021occupant}. These works use structural vibrations from footsteps to identify customer movement, potentially allowing stores to track foot traffic patterns without the need for cameras or RFID.

Additionally, works have also explored audio-induced shelf vibrations to estimate weight changes and identify objects~\cite{zhang2025wevibe, zhang2023vibration}. This approach could be extended to detecting a person’s presence near shelves, helping track product interaction more effectively.

\subparagraph{\textbf{Sensor Fusion for Enhanced Tracking}}
While each sensing method has limitations, combining various sensing methods can improve and offset these issues~\cite{pan2013securitas,han2018smart,han2017sensetribute,han2018smart}. LiDAR, RFID\textbf{,} and vibration sensors can improve tracking reliability. Multimodal approaches have been used in systems like VibSense \cite{liu2017vibsense}, which leverages touch-based vibration sensing on surfaces to identify customer-product interactions.

By integrating LiDAR (depth mapping), RFID (shopper identification), and vibration sensing (movement tracking), an autonomous store can achieve:

\begin{itemize}
    \item High-accuracy people tracking without cameras
    \item Group behavior differentiation
    \item Enhanced theft prevention with privacy compliance
\end{itemize}

\subsubsection{Future Directions \& Improvements}
While multimodal sensing improves tracking accuracy, future research should explore:

\begin{itemize}
    \item Machine learning models to interpret sensor fusion data more effectively.
    \item Energy-efficient tracking systems to reduce the computational cost of multimodal approaches.
    \item Combining real-time inventory tracking with foot traffic analysis for better store optimization.
\end{itemize}

\section{Sensing Approaches}
Autonomous stores rely on a variety of sensing technologies to track inventory, monitor customer behavior, and enhance security. Each sensing approach comes with unique advantages and challenges, making it necessary to explore multi-modal solutions that combine different sensors for improved accuracy and efficiency. This section provides an overview of key sensing modalities, including camera-based vision systems, vibration sensors, RFID tracking, LiDAR, ultrasound, and passive infrared (PIR) sensing, highlighting their roles in inventory management, theft detection, and customer interaction tracking.

\subsection{Camera}
Camera-based sensing is prevalent and is also a core component in autonomous settings. Camera sensors provide an advantage over existing sensors with its extensive coverage range and ability to simultaneously provide rich visual data on multiple items.  Leveraging computer vision, cameras are used to perform tasks such as localization, item identification and classification, and item movement. 

\subsubsection{Localization and Movement}
One of the main applications for cameras is precise localization in diverse environments. Studies \cite{dong2022gaitvibe+,dong4987438robust,purohit2013sugartrail, ruiz2020idiot} have demonstrated how cameras are able to perform location tracking tailored to specific environments. Researchers are able to track the position and orientation of humans or objects using skeletal key points extracted from camera feeds. When combined with multi-view geometry \cite{dusmanu2020multi, heng2019project, cai2019camera} or depth estimation \cite{alam2023review, kim2012multifocusin, brachmann2021visual}, this visual data allows systems to calculate the relative positioning and infer movement trajectories within a known space . In autonomous stores, this system can help track customer activity and detect inventory movement. On top of solely camera-based data, current approaches explore augmenting vision-based sensors with complementary sensors to enhance temporal resolution and address privacy concerns. 

\subsubsection{Item Classification}
Camera-based identification uses computer vision to analyze and interpret video frames. In retail environments, identification systems analyze visual components like dimensions, colors, and shapes using deep learning algorithms \cite{gothai2022design, franco2017grocery}. These systems can distinguish between similar-looking items with different colors schemes, providing a high accuracy rate even in crowded retail environments, but can be further refined with character recognition technology \cite{georgieva2020optical, sterkens2022computer, baldominos2019survey, khan2021machine, fujitake2024dtrocr}. These systems pick up product labels and packaging texts from images much like the text localization techniques humans follow when given a visual data. With character recognition in place, it could provide more robust identification results that effectively spots minimal differences between physically similar retail items.

\subsection{Vibration}

Vibration sensors are widely used for detecting seismic events \cite{deng2013mems, weng2011robust, dean2016distributed, stewart2016unmanned}, intrusions \cite{kuccukbay2017use, ali2015intrusion, hart2016risks, gomery2000fence, wang2021intruder, mukhopadhyay2017detection}, and structural damage \cite{deraemaeker2006vibration, panagiotopoulos2023damage, yang2002damage, farrar2001vibration}.
Furthermore, they have been explored for localization \cite{schloemann2015vibration, poston2015towards, mirshekari2018occupant, poston2016indoor, kashimoto2016floor} and object tracking \cite{pan2016multiple, pan2017surfacevibe, drira2022framework, poston2018toward, pan2018characterizing, drira2019model}, demonstrating their versatility across various sensing applications..

Given their versatility, vibration-based sensing has emerged as a promising approach for inventory tracking, theft detection, and even physiological monitoring in autonomous stores. Unlike vision-based or RFID-based tracking, vibration sensors can provide indirect, contactless detection of customer interactions and product movements, offering a cost-effective alternative for real-time monitoring.

\subsubsection{Vibration for Inventory Monitoring}
One key application of vibration sensing in autonomous stores is detecting product interactions on shelves. \cite{zhang2025wevibe, zhang2023vibration} leverages controlled audio-induced vibrations to monitor weight changes on shelves, and \cite{saidi2023smart} exploits imbalanced load cells to infer product positions, allowing for real-time product detection without requiring dense load sensors. Similarly, other works \cite{pan2017surfacevibe, liu2017vibsense, zhang2023single} demonstrate how surface vibrations can track taps, swipes, and even weight shifts to infer customer interactions.

\subsubsection{Vibration for Theft Detection}
Vibration-based sensing can enhance security in autonomous stores by detecting suspicious movements and potential shoplifting behaviors. \cite{pan2017footprintid} demonstrates how ambient floor vibrations can be used to identify individual footsteps, allowing for real-time customer tracking. Similarly, \cite{codling2024flohr} explores the use of indirect floor vibration sensing for heart rate and presence detection, which could help infer signs of stress or unusual behavior that may indicate theft.

Several studies have investigated the integration of vibration sensors into anti-theft systems, particularly in vehicle security \cite{hossain2017design, liu2013vehicle, mamun2015anti, krishnaprasad2021novel}. Extending such techniques to high-value retail items presents a promising direction for theft prevention. However, existing systems often rely on a combination of multiple sensors, GPS trackers, and alarms, making them costly and complex for retail environments. Additionally, vehicle anti-theft systems typically monitor large, stationary objects, whereas retail products vary in size and weight, making vibration-based detection less straightforward. Furthermore, shoplifting presents a more diverse attack model, as legitimate customer interactions such as picking up, inspecting, or putting items back introduce significant ambiguity, making it difficult to distinguish between theft and normal shopping behavior. Developing a lightweight and cost-efficient vibration-based theft detection system remains an open challenge, requiring further exploration into sensor placement, signal processing, and integration with existing store security infrastructure.

\subsubsection{Vibration For Product Identification}
Recent work in vibration-sensing leverage sophisticated signal processing techniques and sensor fusion to enable fine-grained sensing beyond basic movement detection. These approaches \cite{zhang2023cpa, zhang2023augmenting, chen2016learning, gadre2022milton} leverage structural vibrations generated by products with different physical materials and characteristics. The product movements generate varying vibrational patterns, which is then used as a distinct spectral signature for classification. 

\subsection{RFID}
RFID (radio frequency identification) is a technology that utilizes low cost wirelessly powered tags, which can be detected by a reader system. RFID tags have been incorporated into cashierless store systems \cite{szabo2023systematic, al2013rfid, maulana2021self} for item to customer association in two main ways: exit detection based systems, where tags are scanned when a customer leaves the store, cart based systems, which rely on a RFID scanner inside a shopping cart, or tracking systems, where long range RFID scanners track the position of tags within an area. 
\subsubsection{Exit Based RFID Systems}
Exit based RFID detection systems \cite{rathore2011rfid, melia2013enhancing} have seen varying implementations. A baseline design of an RFID checkout system  \cite{6524348} replaces self checkout barcode scanners with an RFID system, where users place items with attached RFID tags onto an antenna for detection. While items are identified faster than traditional barcodes, this system’s detection accuracy decreases when multiple tags are clustered together. More advanced implementations \cite{bocanegra2020rfgo, busu2011auto} require less effort from customers to use, such as using an exit gate filled with RFID antennas that customers pass while leaving the store. Custom designed RFID readers with multiple antennas are able to scan several items at once for decreased waiting times and improved accuracy. While such approaches deliver a high level of performance, the cost of hardware is a concern for commercial implementations. \cite{piramuthu2014should, wen2010cost}
\subsubsection{RFID Smart Shopping Carts}
Smart shopping carts \cite{sawant2015rfid, agarwal2011rfid, machhirke2017new} are another implementation of RFID for cashierless stores. Common smart shopping cart designs \cite{pradhan2017konark} feature a modified shopping cart with a RFID reader located at the bottom of the basket. A popular technique to ensure that only tags inside the cart are detected is through measuring the variance in RSSI. One additional benefit of measuring RSSI is that it is possible to track items the user may have interacted with outside the cart, which can provide helpful insight to store owners. Another approach \cite{10486478} for item in cart detection is to scan an item’s RFID tag before it can be placed inside the cart. When an item is scanned, a door located on the cart opens, and the item can be inserted. If the user no longer wants the item, it can be scanned again to remove it from the cart’s inventory. While this approach resolves the issue of scanning multiple RFID tags in a small area, it also features lower security and greater mechanical complexity through the use of a door.
\subsubsection{RFID Based Localization}
RFID based position tracking and localization \cite{sanpechuda2008review, li2019review} is another idea that has been explored in cashierless store development. One aspect of localization in a cashierless store is the identification of item positions on shelves. A solution \cite{5751704} is to have items with UHF RFID tags attached placed on a shelf with reference RFID tags, with a centralized RFID reader covering multiple shelves. When a new item is placed near a reference tag, an increase in interference occurs in the reference tag's signal, which gives an estimate of the item’s position on the shelf. Item removal can be detected by comparing the tag’s RSSI to that to the typical variance in RSSI for the environment around the scanner. Room scale localization is another important component of cashierless stores. One idea for this challenge \cite{9219250} is to attach RFID readers to shopping carts to track cart and customer positions. Using wall mounted boards covered in RFID tags with specific IDs, the cart based reader can detect its own position relative to the the tags. Additionally, if the tags are occluded by the shopper, the system is still able to detect the cart’s position, along with that of the shopper. Combinations of shelf and room based systems can be used to track items and shoppers across an entire store.

\subsection{PIR and LiDAR}
{

}

\subsection{Ultrasound}
Ultrasound sensing is a technique that utilizes high-frequency sound waves to detect object movements, track inventory, and enhance security in autonomous stores. Unlike vision-based systems, which can be affected by lighting conditions, or RFID-based approaches that require tag attachment, ultrasound sensing provides a contactless, infrastructure-efficient solution for monitoring product interactions and customer behavior.

Ultrasound sensors emit acoustic waves typically in the range of 20 kHz to MHz, beyond the range of human hearing. These waves propagate through the environment and reflect off objects, with the time-of-flight (ToF) and Doppler shifts of the reflected waves being analyzed to determine object distance, movement velocity, and surface characteristics. The ToF principle measures the delay between wave transmission and reception to calculate object positioning with high accuracy. Meanwhile, the Doppler effect is used to detect motion by evaluating frequency shifts in the reflected signals. More advanced implementations employ phased arrays, which manipulate wavefronts through constructive and destructive interference, enabling the formation of directional beams for improved spatial resolution. Beamforming techniques further enhance detection by filtering out unwanted reflections and focusing on specific areas of interest.

Distributed Acoustic Sensing extends ultrasound sensing principles by utilizing fiber optic cables as continuous acoustic sensors. In DAS systems, an interrogator unit transmits laser pulses through an optical fiber, and minute strain variations caused by acoustic waves induce changes in the backscattered light due to Rayleigh scattering. By analyzing these backscattered signals, DAS enables real-time detection of vibrations, movements, and disturbances along the entire length of the fiber. This technology provides a scalable and cost-effective solution for large-area monitoring without the need for discrete sensor installations. In autonomous store environments, DAS can complement ultrasound-based tracking by detecting structural vibrations associated with foot traffic, shelf interactions, and potential security breaches, further enhancing the robustness of contactless inventory and customer behavior monitoring.

\section{Sensor Positioning and Distance}

\subsection{Optimizing Camera Placement for Comprehensive Coverage} 
The effectiveness of sensor technologies in autonomous retail environments is heavily influenced by their strategic positioning and distance from tracked items. According to Ruiz et al. (2019), the integration of multi-modal sensing in autonomous inventory monitoring relies critically on optimal camera placement to overcome occlusion challenges and achieve comprehensive coverage. Their research demonstrates that cameras positioned at varying heights and angles significantly improve the system's ability to track customer-product interactions while minimizing blind spots. Similarly, Falcão et al. (2020) examined the impact of camera distance on product recognition accuracy in their FAIM framework, finding that multi-angle video collection from various distances improved the system's ability to distinguish between similar-looking products by allowing for more precise measurement of Mahalanobis distance between product class centroids

\subsection{Addressing Visual Domain Adaptation Through Sensor Positioning} 
Visual domain adaptation presents another critical challenge in sensor positioning for autonomous stores. AutoTag research by Ruiz et al. (2019) specifically addresses scenarios where training data is collected under different conditions than those found in actual store environments, including variations in perspective, distance, and lighting. Their findings suggest that adaptive positioning of vision sensors can mitigate these domain shift issues, with optimal results achieved when cameras can capture products from multiple perspectives that closely match those used during training. The research further indicates that combining adjustable-distance cameras with weight sensors creates a more robust system capable of adapting to varying store layouts and product arrangements. This multi-modal approach compensates for distance-related degradation in visual recognition accuracy by correlating weight measurements with visual data.

\subsection{RFID Sensor Distance Calibration for Effective Product Tracking} 
RFID-based tracking systems also demonstrate significant sensitivity to sensor-product distances. The research by Kim et al. (2010) in their SPAMMS framework highlights how the effective range of RFID readers must be carefully calibrated based on store layout and product density. Their system dynamically adjusts tracking parameters based on distance measurements, with reader positioning optimized to maintain consistent tag detection while preventing signal interference between closely packed items. Similarly, IEEE research on RFID robots for retail item location (2025) demonstrates how mobile sensing platforms can overcome fixed-position limitations by dynamically adjusting their distance from shelved products, achieving optimal read rates at specific distance thresholds tailored to different product categories and packaging materials.

\subsection{Hierarchical Sensor Deployment for Customer Interaction Monitoring} 
Distance considerations extend beyond product tracking to customer interaction monitoring. The AIM3S framework developed by Ruiz et al. (2019) employs strategically positioned sensors at varying distances to capture the full spectrum of customer-product interactions. Their research indicates that close-range sensors provide detailed interaction data but suffer from limited coverage, while distant sensors offer broader monitoring capabilities at the expense of granular detection. The optimal configuration involves a hierarchical deployment with distant sensors triggering more detailed analysis from closer sensors when potential interactions are detected. This tiered approach balances comprehensive coverage with detailed interaction tracking while optimizing computational resources required for real-time processing.

\subsection{Adaptive Sensor Positioning for Future Autonomous Stores} 
Emerging research points toward adaptive sensor positioning systems that automatically reconfigure based on store conditions and monitoring requirements. Georgieva and Zhang (2020) propose optical character recognition systems for product identification that incorporate distance-adaptive focusing mechanisms, allowing for consistent performance across varying shelf depths and product placements. Their work demonstrates how autonomous calibration of sensor-to-product distances can significantly enhance recognition accuracy while reducing deployment complexity. Future autonomous store implementations will likely incorporate AI-driven optimization of sensor positioning, automatically adjusting the distance and angle of vision systems, RFID readers, and other sensors to maintain optimal coverage despite changing store layouts, product arrangements, and customer traffic patterns.

\section{Conclusion}
Autonomous stores rely on a combination of sensing technologies to enable cashier-less shopping, real-time inventory tracking, and enhanced security. Each sensing modality—computer vision, RFID, weight sensing, vibration-based detection, LiDAR, and ultrasound—offers unique advantages but also faces limitations in scalability, accuracy, and cost-effectiveness. To address these challenges, multi-modal sensor fusion has emerged as a promising approach, leveraging complementary sensing techniques to improve reliability and reduce deployment costs.

Despite significant advancements, open challenges remain in inventory tracking granularity, theft detection accuracy, and privacy-conscious customer monitoring. Future research should focus on energy-efficient, cost-effective, and privacy-preserving sensing systems that can seamlessly integrate into autonomous retail environments. Continued improvements in AI-driven sensor processing, real-time data fusion, and adaptive sensing techniques will be key to making autonomous stores more scalable and efficient.

\bibliographystyle{ACM-Reference-Format}
\bibliography{references.bib}

\end{document}